\documentstyle[11pt]{article}

\font\tenrm=cmr10

\newcommand{\be}{\begin{equation}}
\newcommand{\ee}{\end{equation}}

\title{\huge Supersymmetry : A new organizing principle for the 
microworld?\footnote{Presented at the Seminar
 on Philosophy of Science, IIT Bombay, February 1998}}

\author{U. A. Yajnik\thanks{e-mail : yajnik@niharika.phy.iitb.ernet.in}\\
{\tenrm Physics Department, Indian Institute of Technology Bombay,}\\
{\tenrm Mumbai 400\thinspace076}}

\date{}

\begin{document}
\maketitle
\begin{abstract}

{A glorious achievement of twentieth century Physics is the
identification of the building blocks of nature. A related triumph 
a deep understanding of the forces of interaction. In particular,
it has been possible to understand the four fundamental forces
of nature in an elegant and unified mathematical framework. The 
paradigm for this unification has been the principle of gauge symmetry. 

A new approach to achieving further unification relies on
Supersymmetry. This is a symmetry which relates particles of integer
spin with those of half-integral spin and also
dictates the allowed interactions among
these particles. However, no data exist
pointing to the presence of this symmetry; it may
indeed be of no relevance to nature.

In this paper the gauge principle and unification attempts 
are briefly reviewed, and then Supersymmetry is introduced.  
The notion of ``broken" symmetry is discussed, which
may explain why the symmetry is difficult to observe 
at presently accessible energies. But a badly broken manifestation
seems to remove much of the original appeal and elegance of
the theory. We here propose two alternative
views of the status of Supersymmetry from fundamental standpoint.
There is a good reason why it is indeed a fundamental symmetry,
despite badly broken. The alternative is that 
it is an organizing principle, somewhat like valence.
This makes it valuable as a signpost to new Physics.}

\end{abstract}


\section{Prologue}
This paper deals with a topic of current interest in Theoretical Physics
and is presented to the forum of historians and philosophers of science.
It assumes familiarity at the popular level with developments
in High Energy Physics and with basic Quantum Physics. In 
rephrasing technical statements, an attempt is made to remain 
close to the truth, albeit selectively. Math is used sparingly 
but some formulae are displayed in the hope that they will give
the reader an opening into more detailed literature. Sections
2 and 3 deal with the paradigm of symmetry as it has come to be 
understood in this century.

Supersymmetry is an elegant symmetry principle, but seems 
to not be operating in nature in its simplest version. 
Section 4 deals with Supersymmetry; the idea, its appeal
and its failings. Section 5 presents two 
alternatives for the metaphysical status of Supersymmetry 
in case it is indeed discovered.

\section{The intangible microworld}
The macroscopic world directly impinges on the senses
and demands systematizing principles.  Presenting itself 
in many different contexts, it also provides ample clues for 
arriving at such principles.
Majority of the phenomena of common experience are correctly
described by Newtonian principles, complemented by laws
governing electromagnetism, hydrodynamics and so on.

The microscopic world is nevertheless present. It is intangible
except for a few tangible and powerful clues. The shape and 
solidity of the world relies on Fermi statistics. Several
phenomenological constants contain Planck's constant
or Avogadro's number. The scientist has had to progressively
become a detective relying on skimpy evidence to pursue the trail
of an elusive and magnificent if unseeable reality. 

Consider an example of clues leading to detection.
Valence was wrested from Chemical phenomena, after much confusion
and controversy. Mendeleev's periodic table, at first based on 
valence, systematized the elements and predicted new ones. The
raison d'etre of the table remained a mystery until the electronic
structure of the atom could be understood. This in turn needed
the Pauli exclusion principle for its explanation; in turn bringing us
to the very heart of microscopic phenomena, the fundamental
indistinguishability of quanta. The message of Quantum Mechanics 
is that only the possible quantum states of a collection of quanta
that are distinct, not the quanta themselves.

\subsection{The metaphysics of insight}
An important paradigm for theoretical progress in this
century has been symmetry principles. This is in contrast
to the development of Electromagnetism which occurred over
about two centuries, during which theory and experiment
progressed step by step, aiding each other. The exploration
of the microworld beginning with radioactivity did not
enjoy such a luxury of wealth of data, nor of easily
constructible and repeatable experiments. 

The developments starting during 1880's and culminating in 1930's
therefore relied on deep insights guided by 
certain metaphysical  assumptions. 
Here and in the following, by metaphysical we shall mean 
principles external to the discipline of physics itself, 
but nevertheless conceived and used by professionals. 
It is the implicit use of such principles that is the subject
of this paper. Also, being external to the discipline itself,
they appropriately form the subject matter of Philosophy of Science.

There are two simple but deep principles used universally
in science. These are, (1) universal applicability of the concepts,
(2) consistency of the epistemy. By the latter we mean
the expectation that existing technical frameworks or
formalisms will apply also to a relatively new domain of phenomena. 
An example of above principles at work is provided by 
the discovery of Bose Statistics. Planck's explanation
of Black Body radiation relied on assumption of absorption
and emission of radiation energy in quanta. Einstein made
this fact into a new concept, that of a photon, and used
it to explain photoelectric effect. This put the photon
on a more general footing. The next step, which took many
years in coming, was taken by  Bose who assumed that the
thermodynamics of photons must be deducible from a
counting of states just as in classical Statistical
Mechanics. This is what we mean by principle (2). There was
one revolutionary new input required however. The counting of
states is based on strict indistinguishability among the photons.

We cite these as examples of insight working in conjunction
with above guiding principles. In the Quantum domain
however, both proved unreliable. Neither the concepts could
be universally applied, nor the epistemy. Mathematically precise 
entities and rules were in some sense the only infallible
guide.  Which concepts would remain robust and what exactly 
these rules were took a long time for its understanding.
Barring possible new phenomena such as
Hawking radiation, Quantum Mechanics as we know today is
consistent and complete but does not cease to evoke
disbelief even in eminent practitioners.

There was however another metaphysical principle which was
emerging as means of guessing ahead. Very loosely it may
be called the principle that the equations must be elegant
and must incorporate a certain symmetry. 
It was based on this principle that
Einstein's theory of Gravitation and Dirac's theory for the 
relativistic electron were accepted by Physics community with
awe and excitement even before they could be completely
established. In its highly evolved form today, it has come 
to be further formalized as a demand for
the existence of precise, mathematically implementable
symmetry principles.

The origins of this metaphysical 
principle go back to the nineteenth century, when Maxwell
achieved an elegant unification of the laws of electromagnetism.
He organized several laws and rules of thumb then known
so that the Electric and Magnetic forces appeared on par with each 
other displaying an uncanny similarity between the two. 
The laws were also stated in the form of mathematical equations
that permitted easy geometrical visualization and were yet
so far reaching in their import and applicability that an 
eminent colleague is reported to have exclaimed,
quoting Goethe, ``was it a god who wrote those lines?"
With hindsight we know that in fact the symmetry they
displayed went much farther than a nineteenth century esthete
could have discerned, for they were the first equations 
to be known which were covariant under Special
Relativistic transformations.

What we are trying to identify as a principle is
actually rather broad and perhaps contains more logically 
distinct positions than one. We shall focus here on a much
more restrictive and precise aspect of this principle.
Specifically we refer to the use of symmetry principles that tend
to restrict theories and introduce economy of phenomenological
parameters in it. We shall refer to it broadly as  Gauge Symmetry.
It started with Einstein trying to formulate 
relativistically consistent laws of Gravitation, and
later also helped to shape the laws of strong and weak interactions.
The gauge symmetry underlying General Relativity on the one
hand and Gauge Field Theories of strong and weak interactions
on the other hand have several technical differences. But
as has been emphasized by Weinberg\cite{weingrco}, they
have an essential similarity to permit being viewed as
manifestations of the same basic principle.
This is the topic of the next section.

\section{Gauge Symmetry}\label{sec:ggesym}

The most common example of a mathematically implementable
symmetry principle in Physics is the idea of 
rotations.~\footnote{This section 
is an abridged version of the author's contribution to
the Seminar on Philosophy of Science, IIT Bombay, 1993.}
We do not expect outcomes of experiments to depend on the
directional orientation of the apparatus. Considered as a
set of operations, the rotations form a mathematical
structure called Group.
One representation of this group is in terms of
matrices, acting on vectors such as the position vector or
the electric field vector.

The Special Relativity principle is a similar principle, in
fact a generalized rotation involving time, such that the
ordinary rotations form a subgroup of this bigger group.
But this notion of symmetry was completely revolutionized
by Einstein in his subsequent work, viz., General
Relativity.  This theory is actually a theory of Gravity,
generalized from its Newtonian version and made consistent
with Special Relativistic rotations.  The prescription of
General Relativity can be summarized in two parts : (1) The
space-time should be treated as curved, like the surface of
a ball. Thus gravitational influences are described by a
set of space-time dependent functions that specify the 
distance and angle measurement prescriptions. In
a curved space these replace the Pythagorian distance law
from point to point.
(2) In a curved space one does not choose a rigid, 
Cartesian system of coordinates, but
any convenient curvilinear coordinates.  So the laws of
physics must be such as to remain invariant under arbitrary
choices of curvilinear coordinates. This translates to
invariance of the laws under rotations that can be different at
different points.  This two part law was called the
Principle of General Covariance.

This theory was a great speculative triumph. At the time of its
invention, there was no evidence for it.  No one had 
suspected that the perihelion precession of Mercury had 
contributions from Relativistic effects,  
requiring fundamental reformulation of Newtonian Gravity.
No other experimental evidence
existed that demanded such a generalization.

In the 1930's the notion of rotational symmetry was extended by
Heisenberg in a very profound way.  It is known that the
strong nuclear force does not depend separately on the
physical state of the proton or that of the neutron.  In
Quantum Mechanics, the physical state of a system is
described by a complex wavefunction denoted $\psi$. Heisenberg's
proposal was that instead of using $\psi_p$ (for proton)
and $\psi_n$ (for neutrons), if we used 
\begin{eqnarray}
\tilde{\psi_p} &= c_1 \psi_p + c_2 \psi_n\cr
&\cr
{\rm and}\qquad
\tilde{\psi_n} &= c_3 \psi_n + c_4 \psi_p
\end{eqnarray}
the physics would remain unchanged.  Here the $c_1, c_2$, $c_3, c_4$
are complex numbers satisfying some constraints.  The relations
above can be thought of as a complex rotation, an abstract
generalization from the case of real vectors.  
This rotation, called an isospin rotation is a
symmetry (although approximate) of the strong nuclear force.

Several decades later, Yang and Mills proposed Gauge Field 
Theories. These were a generalization of isospin symmetry
in much the same way as General Relativity generalized Special
Relativity. Both prescribed the form of the interaction,
although the requirement was stated as a geometrical law.

To summarize, precise mathematical principles were used
as a {\it strategy} to guess at a theory with insufficient
experimental evidence. The theory could well have proved
wrong. This too has happened many times, as for instance
with the original Kaluza-Klein theory. But the success of
the cases in which this approach has worked is spectacular.

\subsection{Broken symmetry} The curious fact about
symmetries is that sometimes they may not be manifest in
the data. This can happen due to two different reasons.
One reason is that the symmetry may be only approximate.
That is, only by ignoring some of the data or by modifying
their values does one see the symmetry principle at work.
For this to be true, the contaminating effects should be
small in a quantitative sense. But this case of
non-manifest symmetry is not as interesting as the next
one. 

It has been found that in some systems, the governing
equations possess a certain symmetry. However, the 
complexity of the interactions drives the system to
solutions that do not reflect the symmetry. This case
is called Spontaneous Breakdown of symmetry. In case of
weak nuclear interactions, one seeks a theory that obeys
gauge invariance, somewhat similar to the two-part
principle of General Covariance. However, gauge invariance
implies masslessness of the mediating particles, whereas
the mediators of the weak force are known to be massive.
The resolution of this paradox lay in realizing that there
could be additional particles, known as Higgs whose complex
dynamics leads to the ground state of the system not
explicitly displaying the gauge symmetry. Under these 
circumstances, the interaction of the gauge particles with 
the Higgs particles makes the former massive.

In the second type of broken symmetry, the symmetry is all
the time present, being made invisible by the particular
state in which the system is available to us.
In this case, guessing the governing equations is difficult 
but symmetry can be used  as a guiding principle.

\section{Supersymmetry}

This brief history prepares us for a description of the
new proposal of Supersymmetry. The origins of the search
for this rather bizarre symmetry to be described lie in
two unrelated motivations. One was a direct one, asking
whether photon and neutrino, the only two particles
known to be massless in 1960's had anything more in common.
Specifically whether they were two manifestations of the
same particle ``species" masquerading as two.
Secondly, it was also 
a search for {\it the most general} type of symmetries
allowed by interactions that respect the basic rotational
and Special Relativistic symmetries of space time.
There was also an enigma in the distinction between
the gauge symmetry of Gravity which involved space-time 
itself and the gauge symmetry of Nuclear forces, which
seemed to operate in an abstract space of wave functions.
The General Covariance of Gravity came to be called an
external gauge symmetry and the Gauge symmetry of the
nuclear forces internal symmetries. The possible kinds of
internal symmetries were soon classified in terms of
the mathematical theory of Lie Groups. It is a rich variety
of possible symmetries. The question was,
what were the most general kinds of {\it external} symmetry
and whether there could be any mixing between external and
internal.

The above question was supposedly answered with some degree
of finality by a so called "No-go" theorem which appeared
in the late sixties. It said that all the possible symmetries
one could possibly have, consistent with Quantum Mechanics
were the ones already known, viz., a variety of internal
symmetries like isospin on the one hand and the 
already known external symmetries, those of the Special
Theory of Relativity (subsuming the old known symmetry of
rotations) on the other hand.  There was nothing new to
be added to the category of external.

There was a loop hole however. In order to understand it,
let us look at a mathematical statement of internal
consistency of several symmetry operations is formulated.
It can be checked by some amount of careful experimentation
that a small amount of $x$-axis rotation followed by a
small amount of $y$-axis rotation, is not the same as
the $y$-rotation first followed by same 
$x$-rotation. 
This can actually be checked by holding up a pen. The
results of the two operations differ by a small $z$-axis 
rotation! This was first put in the form of equations by Hamilton
in mid-nineteenth century. In modern notation one says
\be
L_x L_y - L_y L_x = L_z
\ee
Here $L_x$ stands for the operation of small $x$-axis
rotation. The product $L_xL_y$ has to be read right to left
for its factors. The left hand side is called the
commutator of the two operators.

The case when the commutator of two operators vanishes, is
when the two operations are really independent of each other.
For example small linear motion in $x$ direction is
completely independent of small linear motion in $y$
direction. So they can be taken up in any order, giving the
same result. This fact is expressed by the equation
\be
P_x P_y - P_y P_x = 0
\ee

In the Quantum Theory, formulated in terms of space-time
dependent fields, there is a different kind of
``commutator". It was known since the 1930's that to obtain
a consistent Quantum theory of particles of spin $1/2$, one
must require an {\it anti-commutation} relation. The independence
of quantum field operator at far away points $x$ and $y$ 
has to be expressed by
\be
\psi(x)\psi(y) + \psi(y)\psi(x) = 0\qquad\qquad {\rm(Fermionic)}
\ee
What is unusual about this relation is the plus sign where
minus should be; and this indeed expresses independence.
The correctness of this rule is amply borne out by the
Pauli Exclusion Principle whereby two spin $1/2$ particles
can never occupy the same state. 

The other kind of particles, those with integer spin and
called Bosons, obey a more familiar algebra, where
independence is expressed by
\be
\phi(x)\phi(y) - \phi(y)\phi(x) = 0 \qquad\qquad {\rm(Bosonic)}
\ee
These algebraic operations were however considered special
to the Quantum Fields representing real particles, not to
be confused with operators representing symmetry
operations. The breakthrough against the No-go theorem lay
in realizing that perhaps one could allow ``fermionic"
algebra even between symmetry operations. Thus consider
a linear displacement along two directions which are
independent, and require 
\be
\theta_1\theta_2 +  \theta_2 \theta_1 = 0
\ee
Here $1$ and $2$ are some ``directions" whose meaning
is yet to be clarified, $\theta$ the corresponding
operators. In what way this can be visualized and in
what sense this is an independence are questions not
easy answer. For the moment we take symbols and their
algebra as guides and check for internal consistency of
various operations. Miraculously, it turned out that one
could indeed expand the algebra of the Special Relativistic
generators in this way, provided all the new generators
were fermionic rather than bosonic. This was a possibility
not considered by the authors of the No-go theorem.

\subsection{Superspace}
Here we elaborate a little on the technical idea of
Supersymmetry. Supersymmetry was first formulated as a set of 
operations on Quantum Fields. An interpretation closer to that for
usual Special Relativistic symmetries was later formulated, pioneered
among others by Abdus Salam. To every of the four dimensions $(t, x, y, z)$
there corresponds a superspace dimension, and these are labeled
as $(\theta^1, {\bar \theta}^1, \theta^2, {\bar
\theta}^2)$. They are supposed to obey anticommuting algebra.
The mathematics of classical (non-Quantum) variables of this
kind was known to the mathematicians as Grassmann algebra.
Just as rotations led to a mixing of the axes, a
supersymmetric translation leads to mixing of ordinary and
superspace axes. To give an example, if $\theta^1$ is
shifted to $\theta^1 + \alpha$ then the $x$ coordinate shifts
as 
\be
x \longrightarrow  x - i\alpha {\bar \theta}^2
\ee
This does not mean anything to us according to usual
intuition. But this is how things proceed in gleaning
secrets of the microworld. From abstract operations on
fields, we proceeded to further compatibility with usual
space time picture; the metaphysical principle (2) of section 3.
 Perhaps future knowledge of new
phenomena will help us visualize these operations better.

\subsection{Predictions and extensions}
There are two main results that follow from assuming that 
there is supersymmetry in nature. The first is that for every 
fermion of given mass there is a boson of identical mass 
and vice versa. This means that corresponding to the observed
photon, there must exist a spin half particle which has been
named photino. Similarly corresponding to the electron,
there must exist a spin zero particle which has been named
selectron (abbreviating `scalar electron'). This nomenclature
pattern is followed for all the hypothetical supersymmetric
partners of known particles. 

The problem is, we don't have a single known pair of species 
which may be considered superpartners of each other. It is worth
recalling that the original motivation for searching for
supersymmetry was to identify the almost massless neutrino
as the superpartner of the photon. This however cannot be
true because other quantum numbers as required by the symmetry
principle do not match.

The second and very powerful implication of supersymmetry is that
it subsumes the usual gauge symmetry principle and predicts
{\it all} the possible forms of interactions between the particles.
This is a very desirable and attractive. This was the main benefit
of pursuing symmetry principles. They should help us to guess the
form of the interaction. Since we do not see any superpartners
yet, the confirmation or otherwise of this prediction is in the
future.

There are many other attractive features of supersymmetry from
the theoretical point of view but are more technical. 
And the simplest predictions seem to be unviable.
Does this mean supersymmetry is of no use? The experience
of searching for gauge symmetry for the weak interactions tells us
that we should keep the possibility open that this symmetry too
is not realized in nature in its simple uncomplicated version,
but perhaps exists in a broken form.

The problem of broken supersymmetry is technically more involved
than breaking of the known gauge symmetries. In some sense this is
because the principle is really very strong. It is difficult
to understand how the symmetry breaks. Developing an understanding
of that is itself a theoretical challenge.

Supersymmetry and its possible manifestations constitutes a
subject of extensive scientific investigation
at present. There are several hypothetical models with mechanisms
for breaking supersymmetry and several giant accelerators being constructed
to check these models. In addition, as true with all Elementary
Particle Physics, these models ought also to have left their
imprints in the early Universe. Efforts are also therefore on to
validate or invalidate some of these models based on cosmological 
observations being carried out today. 

\section{Philosophical positions}
The esthetic appeal of the symmetry paradigm lies in the
elegance of the mathematical structure. 
It seems to generalize the ordinary notion of the freedom to
choose the frame of reference. (See sec. \ref{sec:ggesym}).
It also permits mixing or ``rotating" of distinct particle 
species into each other, thus entailing economy in the number 
of particle types or species. On the utilitarian side, the unification
achieved requires fewer coupling constants, since several
are dictated to be identical and others have to be simple
multiples of a basic value. This success however has not been 
unqualified. In the Standard Model of elementary particles
for example, although the advertised benefits are
present, and the coupling constants are fewer, there do remain
a large number of unknown parameters. These arise primarily in
the form of unknown masses of fermionic and scalar species.

Supersymmetry has merited so much attention due to its elegance.
But it does require introducing a large number of new species
or types of particles. Since many of these are fermionic and
scalar, their masses again require a large number of unknown
parameters. The whole picture is further complicated by the need
to have the symmetry broken. The mechanisms that explain this 
breakdown have to rely on more unknown physics, thus invoking 
whole new unknown sectors of the theory. Some of the new particles
are supposed to be unobservable by themselves
and their influences on the observable world are only through
the fact that supersymmetry appears broken. There is supposed
be very little additional observable evidence about them even
in principle. 

How do we view the situation from outside Physics? I submit that
there are two possibilities. One is that Supersymmetry is indeed
a deep new principle. The other is that it is an expedient 
necessary to tide us over till further experiments provide more
clues. The third uninteresting possibility of course remains,
viz., it shares the fate of several other profound speculations 
about nature, beautiful but irrelevant.

\subsection{A principle ...}
There are several technical reasons advanced to support
why it must indeed be a fundamental principle of nature. For example
it is meant to rationalize some of the mystery surrounding the
Electroweak symmetry breaking. We can not enter into this 
discussion. In the spirit of staying close to fundamental facts,
we may yet advance a reason for the same position along the following 
lines. 

There is no known classical analogue to fermions. For
several decades in early Quantum Mechanics they were treated
with awe and mystery. Their wavefunction can distinguish between
$360^{\circ}$ and $720^{\circ}$ rotations. Pauli referred to this
as ``non-classical two-valuedness". However, later developments
have required and guided parallel treatments for bosons and fermions. 
In the path integral approach, fermions could  be 
elegantly included by inventing rules for integration over
fermionic variables. But more importantly, a set of simple consistent
rules are suggested by the formalism itself. This is the first
time that one treats {\it classical} fermionic variables with
impunity and gets the required answers. 

This begins to suggest
that the bias towards bosons as more natural is perhaps purely
classical. As Dirac\cite{dirac}  emphasizes, the early Quantum Mechanics
was developed only for those systems that have a classical analogue.
It was not possible to ``quantize" other systems. But such may
nevertheless exist. Over the decades 
that have elapsed, the only other kind of system that seems to 
require quantization is the fermionic one. (Modulo phenomena we 
still have no reasonable explanation for). Thus the microworld
dictates that we expand our classical notions to include the Grassmann
numbers as well. The notion of Superspace brings further parity
between bosonic and fermionic dimensions. In fact if the fermionic 
dimensions did not exist, bosons would still get away with a special
status. What more elegant framework could we have for understanding
these new dimensions than to require that they are partners in a
very rigorous sense to the bosonic dimensions of common experience.

\subsection{... or an expedient?}
But if the symmetry is so fundamental, why do we not see it in its
pure form? Should the fact that the ideal principle has already been
disproved make us abandon the search for supersymmetry?
This brings us to our second, an inelegant but more pragmatic view.
A usual argument for assuming supersymmetry is as follows.
The newer, higher energy accelerators have to be built with special
purposes in mind, i.e. to search specific energy ranges, and to 
detect particles with expected decay products.
We need guiding principles to channel our searches and supersymmetry
seems to be only elegant guiding principle aside from gauge
symmetry. But we shall go a little bit beyond this position.

Suppose that supersymmetry is indeed discovered. It may manifest
itself in an ugly and highly disproportionate form. Would it be still
worth discovering? The answer is an affirmative. The caveat is
that it may not be due to the pristine principles we advanced.
But supersymmetry may yet act as an ``organizing principle".
By this we mean a rule such as the valence of elements. This
concept allows us to understand the possible compounds the element
can form. Empirical study then also reveals for the same element
several possible valence states. But once discovered in one context,
that valence state can be sought for in other contexts as well.
Valence is not any kind of fundamental principle. But it does 
put strong restrictions on the kind of molecules that can exist.

An ugly manifestation of supersymmetry would be worth having for
the same reason. It may still place a strong restriction on
the kind of fundamental particles that can exist and the qualitative
nature of their interactions. Valence of course is vindicated by
centuries of further developments which make it a direct descendant
of indeed a deep and beautiful principle.
Perhaps the same is true of supersymmetry.

\section{Conclusion}
We argued that the need to understand the microworld has demanded
greater ingenuity from the theoretical physicist in the twentieth
century. The response to this challenge has been in the form of
educated metaphysical principles evolved by the pioneers of the
subject. A prime one of these in retrospect, has been the principle
of symmetry with specific mathematical connotations.
Its great bounty has been the ability to guess at a whole theory
starting from very scarce evidence.
In particular, gauge symmetry has shaped our understanding of
the fundamental forces of nature, beginning with General Theory
of Relativity. Supersymmetry is another such principle, proposed
in advance of any empirical data but with many compulsive reasons
for its correctness. However, even if discovered, the symmetry 
would have to be found in a badly broken form. This seems to detract 
from its original appeal. We have presented a very general reason
why we may expect the symmetry to be fundamental. Yet there is an
alternative possibility that it may be an organizing principle
similar to valence. In either case it is worth stepping up the efforts
to verify its presence or otherwise in nature.


\end{document}